\documentstyle[12pt]{article}

\input{tcilatex}

\begin{document}

\title{Bound - states for truncated Coulomb potentials}
\author{Maen Odeh and Omar Mustafa  \\
Department of Physics, Eastern Mediterranean University\\
G. Magusa, North Cyprus, Mersin 10 - Turkey\\
email: omustafa.as@mozart.emu.edu.tr\\
}
\maketitle

\begin{abstract}
{\small The pseudoperturbative shifted - }$l$ expansion technique PSLET
[16-19] is generalized for states with arbitrary number of nodal zeros.
Bound- states energy eigenvalues for two truncated coulombic potentials are
calculated using PSLET. In contrast with shifted large-N expansion
technique, PSLET results compare excellently with those from direct
numerical integration.\newpage
\end{abstract}

\section{\protect\bigskip Introduction}

\bigskip Attractive truncated Coulomb potentials 
\begin{equation}
V(r)=-\dfrac{1}{(r^{b}+\alpha ^{b})^{1/b}},
\end{equation}
$($ $b=1,2,3.....$ and $\alpha $ is a truncation parameter $)$ are of
special physical interest. They avoid the singularity at $r=0$ ( the crux of
divergence problems ) and serve as models for many interesting physical
phenomena [1-10].For the case $b=1$, Eq.(1) reads 
\begin{equation}
V(r)=-\dfrac{1}{\left( r+\alpha \right) },
\end{equation}
the eminent cutoff Coulomb potential. In quantum-field theory, it has been
suggested that if gravitational interactions of elementary particles are
taken into account, there would be a gravitational cut off of Coulomb
interactions resulting in a finite theory of quantum field. Eq.(2)
represents therefore a nonrelativistic version of this idea. It may,
moreover, be considered as an approximation of the potential of a smeared
charge rather than a point charge.When $b=2$, Eq.(1) implies 
\begin{equation}
V(r)=-\dfrac{1}{(r^{2}+\alpha ^{2})^{1/2}},
\end{equation}
often called the laser-dressed Coulomb potential. A model that has been%
\vspace{0cm} found useful for the study of the spectrum of a laser-dressed
hydrogen-like atoms when exposed to an intensive non-resonant laser-field
[5-10]. It has been shown that under Kramers-Henneberger canonical
transformations the potential of such atoms can be recast into (3) with the
truncation parameter $\alpha $ is related to the strength of the irradiating
laser field [7-9]. The effective potential for scattering by a uniform
spherical charge distribution is well simulated by (3). Moreover, it is very
similar to the modified potential of the nucleus of a muonic atom due to
finiteness in its size [5,11-15].

As the Schr\H{o}dinger equation for neither of the potentials is amenable to
a general analytic solution, one has to retain perturbation techniques or
numerical methods to analyze their bound states. Mehta and Patil [2] had
investigated analytically the s-state energy for the potential (2); however,
no numerical results were obtained . Intensive analysis had been carried out
by De Meyer and Vanden Berghe [3] and Fernandeze [4] on the bound states of
(2). The shifted 1/N expansion technique had been employed to calculate the
energy eigenvalues of potential (3) [6]. David Singh et al [5] have employed
a numerical method to calculate the energy eigenvalues for the potentials
(2) and (3). As such, these potentials are good candidates to be analyzed
through an analytical ( often semi-analytical ) technique to resolve their
underlying physical aspects.

Recently, we have introduced a pseudoperturbative shifted-$l$ ( $l$ is the
angular momentum quantum number ) expansion technique ( PSLET ) to solve for
nodeless states of Schr\={o}dinger equation. It simply consists of using $1/%
\bar{l}$ as a pseudoperturbation parameter, where $\bar{l}=l-\beta $, and $%
\beta $ is a suitable shift. The shift $\beta $ is vital for it removes the
poles that would emerge, at lowest orbital states with $l=0$, in our
proposed expansion below. Our new analytical, often semi-analytical,
methodical proposal PSLET has been successfully applied to quasi -
relativistic harmonic oscillator [16], spiked harmonic oscillator [17],
anharmonic oscillators [18], and two dimensional hydrogenic atom in an
arbitrary magnetic field[19].

Encouraged by its satisfactory performance in handling nodeless states, we
generalize PSLET recipe ( in section 2 ) for states with arbitrary number of
nodal zeros, $n_{r}\geq 0$. In section 3 we apply PSLET to treat the
potentials (2) and (3) and we compare the results obtained by PSLET with the
exact ones. We conclude in section 4.

\section{\protect\bigskip \protect\bigskip Method}

The prescription of our technique starts with the radial part of the
time-independent Schr\H{o}dinger equation in 
h\hskip-.2em\llap{\protect\rule[1.1ex]{.325em}{.1ex}}\hskip.2em%
$=m=1$ units, 
\begin{equation}
\left[ -\frac{1}{2}\frac{d^{2}}{dr^{2}}+\frac{l(l+1)}{2r^{2}}+V(r)\right]
\Psi _{n_{r},l}(r)=E_{n_{r},l}\Psi _{n_{r},l}(r).
\end{equation}
where $l$ is the angular momentum quantum number, $n_{r}=0,1,\cdots $ counts
the nodal zeros and $V(r)$ is an arbitrary spherically symmetric potential
that supports bounded states.

Most textbook perturbation techniques manipulate the potential term to
introduce a perturbating expansion parameter. To the contrary, our method
keeps the potential arbitrary to the condition of being well-behaved. We
invest the centrifugal term to play this role. Thus, with $\bar{l}=l-\beta $
( $\beta $ to be determined in the sequel ), Eq.(4) reads  
\begin{equation}
\left\{ -\frac{1}{2}\frac{d^{2}}{dr^{2}}+\frac{\bar{l}^{2}+(2\beta +1)\bar{l}%
+\beta (\beta +1)}{2r^{2}}+V(r).\right\} \Psi _{n_{r},l}(r)=E_{n_{r},l}\Psi
_{n_{r},l}(r),
\end{equation}
Apparently, the natural limit of Eq.(5) is the large-$l$ limit. At that
limit the centrifugal term dominates over the kinetic energy and the
potential terms. This results in a semiclassical motion of the particle in
an effective potential $V_{eff}=\dfrac{1}{2r^{2}}+\dfrac{1}{Q}V(r)$, where $Q
$ is a constant that scales the potential at large-$l$ limit and is set, for
any specific choice of $l$ and $n_{r},$ equal to $\bar{l}^{2}$ at the end of
calculations[19,20].Hence, the motion is concentrated about the minimum of $%
V_{eff}$, say $r_{0}$. Consequently, a coordinate transformation through 
\begin{equation}
x=\bar{l}^{1/2}(r-r_{o})/r_{o},
\end{equation}
will be in point. It worths mentioning that the scaled coordinates, Eq.(6),
has no effect on the energy eigenvalues, which are coordinates independent.
It just facilitates the calculations of both the energy eigenvalues and
eigenfunctions. Performing the coordinate transformation, Eq.(6), Eq. (5)
reads 
\begin{equation}
\left[ -\frac{1}{2}\frac{d^{2}}{dx^{2}}+\frac{r_{o}^{2}}{\bar{l}}\tilde{V}%
(x(r))\right] \Psi _{n_{r},l}(x)=\frac{r_{o}^{2}}{\bar{l}}E_{n_{r},l}\Psi
_{n_{r},l}(x),
\end{equation}

\begin{equation}
\tilde{V}(x(r))=\frac{\bar{l}^{2}+(2\beta +1)\bar{l}+\beta (\beta +1)}{%
2r_{0}^{2}\left( 1+\dfrac{x}{\sqrt{\bar{l}}}\right) ^{2}}+\frac{\bar{l}^{2}}{%
Q}V(x(r))
\end{equation}
\ \ \ \ \ \ \ \ \ \ \ \ \ \ \ \ \ \ \ \ \ \ \ \ \ \ \ \ \ \ \ \ \ \ \ \ \ \
\ \ \ \ \ \ \ \ \ \ \ \ \ \ \ \ \ \ \ \ \ \ \ \ \ \ \ \ \ \ \ \ \ \ \ \ \ \
\ \ \ \ \ \ \ \ \ \ \ \ \ \ \ \ \ \ \ \ \ \ \ \ \ \ \ \ \ \ \ \ \ \ \ \ \ \
\ \ \ \ \ \ \ \ \ \ \ \ \ \ \ \ \ \ \ \ \ \ \ \ \ \ \ \ \ \ \ \ \ \ \ \ \ \
\ \ \ \ \ \ \ \ \ \ \ \ \ \ \ \ \ \ \ \ \ \ \ \ \ \ \ \ \ \ \ \ \ \ \ \ \ \
\ \ \ \ \ \ \ \ \ \ \ \ \ \ \ \ \ \ \ \ \ \ \ \ \ \ \ \ \ \ \ \ \ \ \ \ \ \
\ \ \ \ \ \ \ \ \ \ \ \ \ \ \ \ \ \ \ \ \ \ \ \ \ \ \ \ \ \ \ \ \ \ \ \ \ \
\ \ \ \ \ \ \ \ \ \ \ \ \ \ \ \ \ \ \ \ \ \ \ \ \ \ \ \ \ \ \ \ \ \ \ \ \ \
\ \ \ \ \ \ \ \ \ \ \ \ \ \ \ \ \ \ \ \ \ \ \ \ \ \ \ \ \ \ \ \ \ \ \ \ \ \
\ \ \ \ \ \ \ \ \ \ \ \ \ \ \ \ \ \ \ \ \ \ \ \ \ \ \ \ \ \ \ \ \ \ \ \ \ \
\ \ \ \ \ \ \ \ \ \ \ \ \ \ \ \ \ \ \ \ \ \ \ \ \ \ \ \ \ \ \ \ \ \ \ \ \ \
\ \ \ \ \ \ \ \ \ \ \ \ \ \ \ \ \ \ \ \ \ \ \ \ \ \ \ \ \ \ \ \ \ \ \ \ \ \
\ \ \ \ \ \ \ \ \ \ \ \ \ \ \ \ \ \ \ \ \ \ \ \ \ \ \ \ \ \ \ \ \ \ \ \ \ \
\ \ \ \ \ \ \ \ \ \ \ \ \ \ \ \ \ \ \ \ \ \ \ \ \ \ \ \ \ \ \ \ \ \ \ \ \ \
\ \ \ \ \ \ \ \ \ \ \ \ \ \ \ \ \ \ \ \ \ 

Expansions about $x=0$, i.e. $r=r_{0}$, yeild 
\begin{equation}
\frac{1}{r_{0}^{2}\left( 1+\dfrac{x}{\sqrt{\bar{l}}}\right) ^{2}}%
=\sum_{n=0}^{\infty }(-1)^{n}\frac{(n+1)}{r_{o}^{2}}x^{n}\bar{l}^{-n/2},
\end{equation}

\begin{equation}
V(x(r))=\sum_{n=0}^{\infty }\left( \frac{d^{n}V(r_{o})}{dr_{o}^{n}}\right) 
\frac{(r_{o}x)^{n}}{n!}\bar{l}^{-n/2}
\end{equation}
Apparently, the expansions in (9) and (10) center the problem at the point $%
r_{0}$ and the derivatives, in effect, contain information not only at $%
r_{0} $ but also at any point on the axis, in accordance with Taylor's
theorem. It is also convenient to expand $E$ as\newline
\begin{equation}
E_{n_{r},l}=\sum_{n=-2}^{\infty }E_{n_{r},l}^{(n)}\bar{l}^{-n}.
\end{equation}
\newline
Equation (7) thus becomes\newline
\begin{equation}
\left[ -\frac{1}{2}\frac{d^{2}}{dx^{2}}+\frac{r_{o}^{2}}{\bar{l}}\tilde{V}%
(x(r))\right] \Psi _{n_{r},l}(x)=r_{0}^{2}\left( \sum_{n=-2}^{\infty
}E_{n_{r},l}^{(n)}\bar{l}^{-(n+1)}\right) \Psi _{n_{r},l}(x),
\end{equation}
\newline
with\newline
\begin{eqnarray}
\frac{r_{o}^{2}}{\bar{l}}\tilde{V}(x(r)) &=&r_{o}^{2}\bar{l}\left[ \frac{1}{%
2r_{o}^{2}}+\frac{V(r_{o})}{Q}\right] +\bar{l}^{1/2}\left[ -x+\frac{%
V^{^{\prime }}(r_{o})r_{o}^{3}x}{Q}\right]  \nonumber \\
&+&\left[ \frac{3}{2}x^{2}+\frac{V^{^{\prime \prime }}(r_{o})r_{o}^{4}x^{2}}{%
2Q}\right] +(2\beta +1)\sum_{n=1}^{\infty }(-1)^{n}\frac{(n+1)}{2}x^{n}\bar{l%
}^{-n/2}  \nonumber \\
&+&r_{o}^{2}\sum_{n=3}^{\infty }\left[ (-1)^{n}\frac{(n+1)}{2r_{o}^{2}}%
x^{n}+\left( \frac{d^{n}V(r_{o})}{dr_{o}^{n}}\right) \frac{(r_{o}x)^{n}}{n!Q}%
\right] \bar{l}^{-(n-2)/2}  \nonumber \\
&+&\beta (\beta +1)\sum_{n=0}^{\infty }(-1)^{n}\frac{(n+1)}{2}x^{n}\bar{l}%
^{-(n+2)/2}+\frac{(2\beta +1)}{2},
\end{eqnarray}
\newline
where the prime of $V(r_{o})$ denotes derivative with respect to $r_{o}$.
Equation (12) is exactly of the type of Schr\"{o}dinger equation for one -
dimensional anharmonic oscillator\newline
\begin{equation}
\left[ -\frac{1}{2}\frac{d^{2}}{dx^{2}}+\frac{1}{2}\Omega ^{2}x^{2}+\xi
_{o}+P(x)\right] X_{n_{r}}(x)=\lambda _{n_{r}}X_{n_{r}}(x),
\end{equation}
\newline
where $P(x)$ is a perturbation - like term and $\xi _{o}$ is a constant. A
simple comparison between Eqs.(12), (13) and (14) implies\newline
\begin{equation}
\xi _{o}=\bar{l}\left[ \frac{1}{2}+\frac{r_{o}^{2}V(r_{o})}{Q}\right] +\frac{%
2\beta +1}{2}+\frac{\beta (\beta +1)}{2\bar{l}},
\end{equation}
\newline
\begin{eqnarray}
\lambda _{n_{r}} &=&\bar{l}\left[ \frac{1}{2}+\frac{r_{o}^{2}V(r_{o})}{Q}%
\right] +\left[ \frac{2\beta +1}{2}+(n_{r}+\frac{1}{2})\Omega \right] 
\nonumber \\
&+&\frac{1}{\bar{l}}\left[ \frac{\beta (\beta +1)}{2}+\lambda _{n_{r}}^{(0)}%
\right] +\sum_{n=2}^{\infty }\lambda _{n_{r}}^{(n-1)}\bar{l}^{-n},
\end{eqnarray}
and\newline
\begin{equation}
\lambda _{n_{r}}=r_{o}^{2}\sum_{n=-2}^{\infty }E_{n_{r},l}^{(n)}\bar{l}%
^{-(n+1)},
\end{equation}
\newline
Equations (15) and (16) yield\newline
\begin{equation}
E_{n_{r},l}^{(-2)}=\frac{1}{2r_{o}^{2}}+\frac{V(r_{o})}{Q}
\end{equation}
\newline
\begin{equation}
E_{n_{r},l}^{(-1)}=\frac{1}{r_{o}^{2}}\left[ \frac{2\beta +1}{2}+(n_{r}+%
\frac{1}{2})\Omega \right]
\end{equation}
\newline
\begin{equation}
E_{n_{r},l}^{(0)}=\frac{1}{r_{o}^{2}}\left[ \frac{\beta (\beta +1)}{2}%
+\lambda _{n_{r}}^{(0)}\right]
\end{equation}
\newline
\begin{equation}
E_{n_{r},l}^{(n)}=\lambda _{n_{r}}^{(n)}/r_{o}^{2}~~;~~~~n\geq 1.
\end{equation}
\newline
Here $r_{o}$ is chosen to minimize $E_{n_{r},l}^{(-2)}$, i. e.\newline
\begin{equation}
\frac{dE_{n_{r},l}^{(-2)}}{dr_{o}}=0~~~~and~~~~\frac{d^{2}E_{n_{r},l}^{(-2)}%
}{dr_{o}^{2}}>0,
\end{equation}
\newline
which in turn gives, with $\bar{l}=\sqrt{Q}$,\newline
\begin{equation}
l-\beta =\sqrt{r_{o}^{3}V^{^{\prime }}(r_{o})}.
\end{equation}
\newline
Consequently, the second term in Eq.(13) vanishes and the first term adds a
constant to the energy eigenvalues. It should be noted that energy term $%
\bar{l}^{2}E_{n_{r},l}^{(-2)}$ is the energy of a particle moving under the
effect of $V_{eff}$. Hence, it is , roughly, the energy of a classical
particle with angular momentum $L_{z}$=$\bar{l}$ executing circular motion
of radius $r_{o}$ in the potential $V(r_{o})$. This term thus identifies the
leading - order approximation, to all eigenvalues, as a classical
approximation and the higher - order corrections as quantum fluctuations
around the minimum $r_{o}$, organized in inverse powers of $\bar{l}$.

The next leading correction to the energy series, $\bar{l}E_{n_{r},l}^{(-1)}$%
, consists of a constant term and the exact eigenvalues of the unperturbed
harmonic oscillator potential $\Omega ^{2}x^{2}/2$. The shifting parameter $%
\beta $ is determined by choosing $\bar{l}E_{n_{r},l}^{(-1)}$=0. This choice
is physically motivated. It requires not only the agreements between PSLET
eigenvalues and the exact known ones for the harmonic oscillator and Coulomb
potentials but also between the eigenfunctions as well. Hence\newline
\begin{equation}
\beta =-\left[ \frac{1}{2}+(n_{r}+\frac{1}{2})\Omega \right] ,
\end{equation}
\newline
where\newline
\begin{equation}
\Omega =\sqrt{3+\frac{q_{o}V^{^{\prime \prime }}(r_{o})}{V^{^{\prime
}}(r_{o})}}.
\end{equation}
Equation (13) thus becomes\newline
\begin{equation}
\frac{r_{o}^{2}}{\bar{l}}\tilde{V}(x(r))=r_{o}^{2}\bar{l}\left[ \frac{1}{%
2r_{o}^{2}}+\frac{V(r_{o})}{Q}\right] +\sum_{n=0}^{\infty }v^{(n)}(x)\bar{l}%
^{-n/2},
\end{equation}
\newline
where\newline
\begin{equation}
v^{(0)}(x)=\frac{1}{2}\Omega ^{2}x^{2}+\frac{2\beta +1}{2},
\end{equation}
\newline
\begin{equation}
v^{(1)}(x)=-(2\beta +1)x-2x^{3}+\frac{r_{o}^{5}V^{^{\prime \prime \prime
}}(r_{o})}{6Q}x^{3},
\end{equation}
\newline
and for $n\geq 2$\newline
\begin{eqnarray}
v^{(n)}(x) &=&(-1)^{n}(2\beta +1)\frac{(n+1)}{2}x^{n}+(-1)^{n}\frac{\beta
(\beta +1)}{2}(n-1)x^{(n-2)}  \nonumber \\
&&+B_{n}x^{n+2},
\end{eqnarray}

\begin{equation}
B_{n}=(-1)^{n}\frac{(n+3)}{2}+\frac{r_{o}^{(n+4)}}{Q(n+2)!}\frac{%
d^{n+2}V(r_{o})}{dr_{o}^{n+2}}
\end{equation}
\newline
Equation (12) thus becomes\newline
\begin{eqnarray}
&&\left[ -\frac{1}{2}\frac{d^{2}}{dx^{2}}+\sum_{n=0}^{\infty }v^{(n)}\bar{l}%
^{-n/2}\right] \Psi _{n_{r},l}(x)=r_{o}^{2}\left[ \sum_{n=1}^{\infty
}E_{n_{r},l}^{(n-1)}\bar{l}^{-n}\right] \Psi _{n_{r},l}(x)  \nonumber \\
&&
\end{eqnarray}
Up to this point, one would conclude that the above procedure is nothing but
an animation of the eminent shifted large-N expansion ( SLNT )[19,20].
However, because of the limited capabilities of SLNT in handling large
-order corrections via the standard Rayleigh-Schr\H{o}dinger perturbation
theory, only low-order corrections have been reported, sacrificing in effect
its preciseness. Therefore, one should seek for an alternative and proceed
by setting the wave functions with any number of nodes as

\begin{equation}
\Psi _{n_{r},l}(x)=F_{n_{r},l}(x)\exp (U_{n_{r},l}(x))
\end{equation}
Eq.(31) is readily transformed into the following Riccati type equation 
\begin{equation}
\begin{array}{l}
-\dfrac{1}{2}\left[ F_{n_{r},l}^{^{\prime \prime
}}(x)+2F_{n_{r},l}^{^{\prime }}(x)U_{n_{r},l}^{^{^{\prime }}}(x)\right]
+F_{n_{r},l}(x){\Huge \{}-\dfrac{1}{2}{\Huge [}U_{n_{r},l}^{^{^{\prime
\prime }}}(x)+(U_{n_{r},l}^{^{\prime }}(x))^{2}{\Huge ]} \\ 
\\ 
+\dfrac{1}{2}(2\beta +1)+\dfrac{1}{2}\Omega ^{2}x^{2}+\sum_{n=1}^{\infty
}v^{(n)}(x)\bar{l}^{-n/2}{\Huge \}}=r_{0}^{2}F_{n_{r},l}(x)\sum_{n=1}^{%
\infty }E_{n_{r},l}^{(n-1)}\bar{l}^{-n}
\end{array}
\end{equation}
where primes denotes derivatives with respect to $x$.It is evident that (33)
admits solutions of the form

\begin{equation}
F_{n_{r},l}(x)=x^{n_{r}}+\sum_{n=0}^{\infty
}\sum_{p=0}^{n_{r}-1}a_{p,n_{r}}^{(n)}x^{p}\bar{l}^{-n/2}
\end{equation}

\begin{equation}
U_{n_{r},l}^{^{\prime }}(x)=\sum_{n=0}^{\infty }U_{n_{r}}^{(n)}(x)\bar{l}%
^{-n/2}+\sum_{n=0}^{\infty }G_{n_{r}}^{(n)}(x)\bar{l}^{-(n+1)/2},
\end{equation}
\newline
where\newline
\begin{equation}
U_{n_{r}}^{(n)}(x)=%
\sum_{m=0}^{n+1}D_{m,n,n_{r}}x^{2m-1}~~~~;~~~D_{0,n,n_{r}}=0,
\end{equation}
\newline
\begin{equation}
G_{n_{r}}^{(n)}(x)=\sum_{m=0}^{n+1}C_{m,n,n_{r}}x^{2m}.
\end{equation}
Clearly the nodal zeros of the wavefunctions are taken care of by $%
F_{n_{r},l}(x)$. For nodeless states,

\begin{equation}
F_{0,l}(x)=1
\end{equation}
which reduces (33) to the problem described in our previous work for
nodeless states[16-19]. To illustrate how our method works for nodal states,
we will treate the one-node state. Upon substituting equations (34)-(37)
with $n_{r}=1$ into (33), it reads 
\begin{equation}
\begin{array}{l}
F_{1,l}(x){\Huge [}-\dfrac{1}{2}\sum_{n=0}^{\infty }\left(
U_{1}^{(n)^{\prime }}(x)\bar{l}^{-n/2}+G_{1}^{(n)^{\prime }}(x)\bar{l}%
^{-(n+1)/2}\right)  \\ 
-\dfrac{1}{2}\sum_{n=0}^{\infty }\sum_{m=0}^{n}{\Huge (}%
U_{1}^{(m)}(x)U_{1}^{(n-m)}(x)\bar{l}^{-n/2}+G_{1}^{(m)}(x)G_{1}^{(n-m)}(x)%
\bar{l}^{-(n+2)/2} \\ 
+2U_{1}^{(m)}(x)G_{1}^{(n-m)}(x)\bar{l}^{-(n+1)/2}{\Huge )}%
+\sum_{n=0}^{\infty }v^{(n)}(x)\bar{l}^{-n/2}-r_{0}^{2}\sum_{n=1}^{\infty
}E_{n_{r},l}^{(n-1)}\bar{l}^{-n}{\Huge ]} \\ 
-F_{1,l}^{^{\prime }}(x)\left( \sum_{n=0}^{\infty }U_{n_{r}}^{(n)}(x)\bar{l}%
^{-n/2}+\sum_{n=0}^{\infty }G_{n_{r}}^{(n)}(x)\bar{l}^{-(n+1)/2}\right) -%
\dfrac{1}{2}F_{1,l}^{^{\prime \prime }}(x)=0
\end{array}
\end{equation}
The solution of (39) then follows from the uniqueness of power series
representation . Therefore equating the coefficients of same powers of $\bar{%
l}$ and $x$ respectively, (of course, the other way around works equally
well), one gets 
\begin{equation}
D_{1,0,1}=-\omega ,\;a_{0,1}^{(0)}=0,\;U_{1}^{(0)}(x)=-\omega x,
\end{equation}

\begin{equation}
C_{1,0,1}=-\dfrac{B_{1}}{\omega },\;a_{0,1}^{(1)}=-\dfrac{C_{1,0,1}}{\omega }%
,
\end{equation}

\begin{equation}
C_{0,0,1}=\dfrac{1}{\omega }(2C_{1,0,1}+2\beta +1),
\end{equation}

\begin{equation}
D_{2,2,1}=\dfrac{1}{\omega }\left( \dfrac{C_{0,0,1}^{2}}{\omega }%
-B_{2}\right)
\end{equation}

\begin{equation}
D_{1,2,1}=\dfrac{1}{\omega }\left( \dfrac{5}{2}D_{2,2,1}+C_{0,0,1}C_{1,0,1}-%
\dfrac{3}{2}(2\beta +1),\right)
\end{equation}

\begin{equation}
E_{1,l}^{(0)}=\dfrac{1}{r_{0}^{2}}\left( \beta (\beta
+1)+a_{0,1}^{(1)}C_{1,0,1}-\dfrac{3D_{1,2,1}}{2}-\dfrac{C_{0,0,1}^{2}}{2}%
\right)
\end{equation}
.... and so on. Obviously, one can calculate the energy eigenvalue and
eigenfunctions from the knowledge of $C_{m,n,n_{r}},D_{m,n,n_{r}}$ and $%
a_{p,n_{r}}^{(n)}$ in a hierarchical manner. Nevertheless, the procedure
just described is suitable for a software package such as MAPLE to determine
the energy eigenvalue and eigenfunction up to any order of the
pseudoperturbation series.

Although the energy series Eq.(11), could appear divergent, or at best,
asymptotic for small $\bar{l},$ one can still calculate the eigen series to
a very good accuracy by performing the sophisticated [N,M] Pade'
approximation[22],

\bigskip 
\begin{equation}
P_{N}^{M}(1/\bar{l})=\dfrac{\left( P_{0}+P_{1}/\bar{l}+\cdot \cdot \cdot
+P_{M}/\bar{l}^{N}\right) }{\left( 1+q_{1}/\bar{l}+\cdot \cdot \cdot +q_{N}/%
\bar{l}^{M}\right) }
\end{equation}
to the energy series, Eq(11). The energy series, Eq(11), is calculated up to 
$E_{n_{r},l}^{(8)}/\bar{l}^{8}$ by 
\begin{equation}
E_{n_{r},l}=\bar{l}^{2}E_{n_{r},l}^{(-2)}+E_{n_{r},l}^{(0)}+\cdot \cdot
\cdot +E_{n_{r},l}^{(8)}/\bar{l}^{8}+O(1/\bar{l}^{9}),
\end{equation}
and with the $P_{4}^{4}(1/\bar{l})$ Pade' approximant it becomes 
\begin{equation}
E_{n_{r},l}[4,4]=\bar{l}^{2}E_{n_{r},l}^{(-2)}+P_{4}^{4}(1/\bar{l}).
\end{equation}
Our technique is therefore well prescribed.

\section{Truncated Coulomb potentials}

\bigskip In this section we consider the truncated Coulomb potentials,
Eqs.(2) and (3).

Substituting Eq.(2) into Eq.(25), one gets 
\begin{equation}
\Omega =\sqrt{\dfrac{r_{0}+3\alpha }{r_{0}+\alpha }}
\end{equation}
Eq.(23) together with Eq.(24) give then 
\begin{equation}
l+\dfrac{1}{2}\left[ 1+(2n_{r}+1)\sqrt{\dfrac{r_{0}+3\alpha }{r_{0}+\alpha }}%
\right] =\dfrac{r_{0}^{3/2}}{(r_{0}+\alpha )}
\end{equation}
When substituting Eq.(3) into Eq.(25), one gets 
\begin{equation}
\Omega =\sqrt{\dfrac{r_{0}^{2}+4\alpha ^{2}}{r_{0}^{2}+\alpha ^{2}}}
\end{equation}
Eq.(23) in turn gives 
\begin{equation}
l+\dfrac{1}{2}\left[ 1+(2n_{r}+1)\sqrt{\dfrac{r_{0}^{2}+4\alpha ^{2}}{%
r_{0}^{2}+\alpha ^{2}}}\right] =\dfrac{r_{0}^{2}}{(r_{0}+\alpha )^{3/4}}
\end{equation}
Eqs.(50) and (52) are explicit equations in $r_{0}$. Clearly, closed - form
solutions of Eqs.(49)-(51) for $r_{0}$ are hard to be found ( which is often
the case ). Thus, we use numerical methods to resolve the issue ( hence the
notion that PSLET is often semianalytical ). Once $r_{0}$ is determined the
coefficients $C_{m,n,n_{r}},D_{m,n,n_{r}},a_{p,n_{r}}^{(n)}$ are determined
in a sequential manner. Hence the eigenvalues, Eq.(47), and eigenfunctions
Eqs.(34)-(37), are calculated in the same batch for each value of $l,n_{r},$
and $\alpha $.

Tables 1 and 2 show PSLET results, Eq.(47), $E_{PSLET}$, for $1s$, $2s$, $2p$%
, $3p$, $3d$, $4d$ and $4f$ eigenstates of Eq.(1), covering week,
intermadiate and strong ranges of the truncation parameter $\alpha $. In
addition, we display the Pade' approximants, Eq. (48), $E[4,4]$, and the
exact results, $E_{exact}$, obtained by direct numerical integration [5].
Comparing $E_{PSLET}$ with $E_{exact}$, one notices the underlying relation
between the accuracy of PSLET results and $l$, $n_{r}$ and the truncation
parameter, $\alpha $. The accuracy of PSLET results increases with
increasing $l$ and/or $n_{r}$ (\ See figure (2) ). This is in accordance
with our choice for the expansion parameter in Eq.(47), as $1/\bar{l}$ gets
smaller as $l$ and/or $n_{r}$ increases. PSLET results show a good
converging trend to the exact values as the truncation parameter gets larger
( See figure (1) ). In the strong range of $\alpha $, PSLET results are
almost exact. To resume the eigenenergy \ \ series, Eq. (47), Pade'
approximant is calculated.  Moreover, the stability noticed in different
order Pade' approximants show that the results are accurate up to ten digits
at strong range of $\alpha $, to the contrary of the of the results obtained
in [5] where the accuracy is recorded up to eight digits in the same range.
\ \ \ \ \ \ \ \ \ \ \ \ \ \ \ \ \ \ \ \ \ \ \ \ \ \ \ \ \ \ \ \ \ \ \ \ \ \
\ \ \ \ \ \ \ \ \ \ \ \ \ \ \ \ \ \ \ \ \ \ \ \ \ \ \ \ \ \ \ \ \ \ \ \ \ \
\ \ \ \ \ \ \ \ \ \ \ \ \ \ \ \ \ \ \ \ \ \ \ \ \ \ \ \ \ \ \ \ \ \ \ \ \ \
\ \ \ \ \ \ \ \ \ \ \ \ \ \ \ \ \ \ \ \ \ \ \ \ \ \ \ \ \ \ \ \ \ \ \ \ \ \
\ \ \ \ \ \ \ \ \ \ \ \ \ \ \ \ \ \ \ \ \ \ \ \ \ \ \ \ \ \ \ \ \ \ \ \ \ \
\ \ \ \ \ \ \ \ \ \ \ \ \ \ \ \ \ \ \ \ \ \ \ \ \ \ \ \ \ \ \ \ \ \ \ \ \ \
\ \ \ \ \ \ \ \ \ \ \ \ \ \ \ \ \ \ \ \ \ \ \ \ \ \ \ \ \ \ \ \ \ \ \ \ \ \
\ \ \ \ \ \ \ \ \ \ \ \ \ \ \ \ \ \ \ \ \ \ \ \ \ \ \ \ \ \ \ \ \ \ \ \ \ \
\ \ \ \ \ \ \ \ \ \ \ \ \ \ \ \ \ \ \ \ \ \ \ \ \ \ \ \ \ \ \ \ \ \ \ \ \ \
\ \ \ \ \ \ \ \ \ \ \ \ \ \ \ \ \ \ \ \ \ \ \ \ \ \ \ \ \ \ \ \ \ \ \ \ \ \
\ \ \ \ \ \ \ \ \ \ \ \ \ \ \ \ \ \ \ \ \ \ \ \ \ \ \ \ \ \ \ \ \ \ \ \ \ \
\ \ \ \ \ \ \ \ \ \ \ \ \ \ \ \ \ \ \ \ \ \ \ \ \ \ \ \ \ \ \ \ \ \ \ \ \ \
\ \ \ \ \ \ \ \ \ \ \ \ \ \ \ \ \ \ \ \ \ \ \ \ \ \ \ \ \ \ \ \ \ \ \ \ \ \
\ \ \ \ \ \ \ \ \ \ \ \ \ \ \ \ \ \ \ \ \ \ \ \ \ \ \ \ \ \ \ \ \ \ \ \ \ \
\ \ \ \ \ \ \ \ \ \ \ \ \ \ \ \ \ \ \ \ \ \ \ \ \ \ \ \ \ \ \ \ \ \ \ \ \ \
\ \ \ \ \ \ \ \ \ \ \ \ \ \ \ \ \ \ \ \ \ \ \ \ \ \ \ \ \ \ \ \ \ \ \ \ \ \
\ \ \ \ \ \ \ \ \ \ \ \ \ \ \ \ \ \ \ \ \ \ \ \ \ \ \ \ \ \ \ \ \ \ \ \ \ \
\ \ \ \ \ \ \ \ \ \ \ \ \ 

Table 3 displays PSLET results, $E_{PSLET}$, for some excited states of
Eq.(3) \ along with the Pade' approximants, $E[4,4],$ of $E_{PSLET}$, the
results obtained by SLNT[6], $E_{SLNT}$ \ and the exact results, $E_{exact}$%
, obtained by direct numerical integration [5]. Apparently, the accuracy of
our results for this potential have similar behavior as in the previous
case. When compared with the results obtained by SLNT, our results show a
better agreement with the exact ones. The difficulty in calculating high
order corrections in SLNT through Rayleigh-Schr\H{o}dinger perturbation
theory results in loss in accuracy. PSLET makes it amenable to calculate
high order corrections which improves the accuracy.

Moreover, one notices that the results of PSLET are more accurate in the
case of the truncated coulomb potential than that of the laser-dressed one.
The reason behind this is that the truncated potential is more coulombic in
nature which makes PSLET nearer to the exact results.

\section{Concluding remarks}

We have presented a generalization of our pseudoperturbative shifted - $l$
expansion technique PSLET [16-19] to treat states with arbitrary number of
nodal zeros, $n_{r}\geq 0.$ Two truncated coulombic potentials have been
treated via PSLET and very accurate energy eigenvalues are obtained.

The outstanding features of the attendant PSLET are in order.

It avoids troublesome questions as those pertaining to the nature of small
parameter expansions, the trend of convergence to the exact numerical values
( marked in tables 1-3 ), the utility in calculating the eigenvalues and
eigenfunctions in one batch to sufficiently higher orders, and the
applicability to a wide range of potentials. Moreover, beyond its promise as
being quite handy, on computational and practical methodical sides, it
offers a useful perturbation prescription where the zeroth- order
approximation inherits a substantial amount of the total energy.

Finally, the scope of PSLET applicability extends beyond the present
truncated coulombic potentials. It could be applied to angular momentum
states of multi - electron atoms [23-25], relativistic and non relativistic
quark - antiquark models [26], etc. We believe that the feature of our
method in determining expressions for the bound-state wavefunctions makes it
possible to describe electron transitions and multiphoton emission occurring
in atomic systems in the presence of an intense laser field. It is therefore
reasonable to re-examine such phenomena in the frame work of PSLET.

{\bf Table1.} Bound-state energies, in 
h\hskip-.2em\llap{\protect\rule[1.1ex]{.325em}{.1ex}}\hskip.2em%
$=m=1$ units, of the potential $V(r)=-\dfrac{1}{(r+\alpha )}$ for the $1s$, $%
2s$, $2p$ and $3p$ states. $E_{PSLET}$ represents PSLET results, Eq (47), $%
E_{44}$ is the [4,4] Pade' approximant, Eq. (48), and $E_{exact}$ from
direct numerical integration [5].

\bigskip

\begin{tabular}{|c|c|c|c|c|c|}
\hline
& $\alpha $ & $1s$ & $2s$ & $2p$ & $3p$ \\ \hline
\multicolumn{1}{|l|}{$-E_{PSLET}$} &  & $0.387357746$ & $0.109481497$ & $%
0.117535370$ & $0.053309085$ \\ 
\multicolumn{1}{|l|}{$-E_{44}$} & $0.1$ & $0.387922157$ & $0.109145059$ & $%
0.117535388$ & $0.053309210$ \\ 
\multicolumn{1}{|l|}{$-E_{exact}$} &  & \multicolumn{1}{|l|}{$0.38754365$} & 
\multicolumn{1}{|l|}{$0.10950805$} & \multicolumn{1}{|l|}{$0.11753535$} & 
\multicolumn{1}{|l|}{$0.05330930$} \\ \hline\hline
&  & $0.180406651$ & $0.069577091$ & $0.082862488$ & $0.041787781$ \\ 
& $1.0$ & $0.180368972$ & $0.069581801$ & $0.082862452$ & $0.041787655$ \\ 
&  & \multicolumn{1}{|l|}{$0.18036705$} & \multicolumn{1}{|l|}{$0.06958066$}
& \multicolumn{1}{|l|}{$0.08286242$} & \multicolumn{1}{|l|}{$0.04178766$} \\ 
\hline\hline
&  & $0.043439053$ & $0.024810349$ & $0.029446519$ & $0.018748152$ \\ 
& $10$ & $0.043438645$ & $0.024810342$ & $0.029446516$ & $0.018748153$ \\ 
&  & \multicolumn{1}{|l|}{$0.04343872$} & \multicolumn{1}{|l|}{$0.02481036$}
& \multicolumn{1}{|l|}{$0.02944652$} & \multicolumn{1}{|l|}{$0.01879815$} \\ 
\hline\hline
&  & $0.012194732$ & $0.008579236$ & $0.009717589$ & $0.007237436$ \\ 
& $50$ & $0.012194683$ & $0.008579235$ & $0.009717588$ & $0.007237436$ \\ 
&  & \multicolumn{1}{|l|}{$0.01219469$} & \multicolumn{1}{|l|}{$0.00857924$}
& \multicolumn{1}{|l|}{$0.00971759$} & \multicolumn{1}{|l|}{$0.00723744$} \\ 
\hline\hline
&  & $0.003653176$ & $0.002907080$ & $0.003169533$ & $0.002608781$ \\ 
& $200$ & $0.003653168$ & $0.002907080$ & $0.003169533$ & $0.002608781$ \\ 
&  & \multicolumn{1}{|l|}{$0.00365317$} & \multicolumn{1}{|l|}{$0.00290708$}
& \multicolumn{1}{|l|}{$0.00316953$} & \multicolumn{1}{|l|}{$0.00260878$} \\ 
\hline
\end{tabular}
\newpage

\bigskip

{\bf Table2.} Bound-state energies, in 
h\hskip-.2em\llap{\protect\rule[1.1ex]{.325em}{.1ex}}\hskip.2em%
$=m=1$ units, of the potential $V(r)=-\dfrac{1}{(r+\alpha )}$ for the $3d$, $%
4d$ and $4f$ states. $E_{PSLET}$ represents PSLET results, Eq (47), $E_{44}$
is the [4,4] Pade' approximant, Eq. (48), and $E_{exact}$ from direct
numerical integration [5].

\bigskip

\begin{tabular}{|c|c|c|c|c|}
\hline
& $\alpha $ & $3d$ & $4d$ & $4f$ \\ \hline
\multicolumn{1}{|l|}{$-E_{PSLET}$} &  & $0.054136568$ & $0.030648450$ & $%
0.030813599$ \\ 
\multicolumn{1}{|l|}{$-E_{44}$} & $0.1$ & $0.054136568$ & $0.030648449$ & $%
0.030813599$ \\ 
\multicolumn{1}{|l|}{$-E_{exact}$} &  & \multicolumn{1}{|l|}{$0.05413657$} & 
\multicolumn{1}{|l|}{$0.03064845$} & \multicolumn{1}{|l|}{$0.03081360$} \\ 
\hline\hline
&  & $0.045010006$ & $0.026625065$ & $0.027588160$ \\ 
& $1.0$ & $0.045010007$ & $0.026625059$ & $0.027588160$ \\ 
&  & \multicolumn{1}{|l|}{$0.04501001$} & \multicolumn{1}{|l|}{$0.02662506$}
& \multicolumn{1}{|l|}{$0.02758816$} \\ \hline\hline
&  & $0.021024302$ & $0.014373461$ & $0.015576600$ \\ 
& $10$ & $0.021024302$ & $0.014373461$ & $0.015576600$ \\ 
&  & \multicolumn{1}{|l|}{$0.02102430$} & \multicolumn{1}{|l|}{$0.01437346$}
& \multicolumn{1}{|l|}{$0.01557660$} \\ \hline\hline
&  & $0.007962796$ & $0.006153630$ & $0.006643882$ \\ 
& $50$ & $0.007962796$ & $0.006153630$ & $0.006643882$ \\ 
&  & \multicolumn{1}{|l|}{$0.00796280$} & \multicolumn{1}{|l|}{$0.00615363$}
& \multicolumn{1}{|l|}{$0.00664388$} \\ \hline\hline
&  & $0.002798562$ & $0.002353647$ & $0.002498272$ \\ 
& $200$ & $0.002798562$ & $0.002353647$ & $0.002498272$ \\ 
&  & \multicolumn{1}{|l|}{$0.00279856$} & \multicolumn{1}{|l|}{$0.00235365$}
& \multicolumn{1}{|l|}{$0.00249827$} \\ \hline
\end{tabular}

\bigskip \newpage

{\bf Table3.} Bound-state energies, in 
h\hskip-.2em\llap{\protect\rule[1.1ex]{.325em}{.1ex}}\hskip.2em%
$=m=1$ units, of the potential $V(r)=-\dfrac{1}{(r^{2}+\alpha ^{2})^{1/2}}$
for the $3d$, $4d$ and $4f$ states. $E_{PSLET}$ represents PSLET results, Eq
(47), $E_{44}$ is the [4,4] Pade' approximant, Eq. (48), $E_{SLNT}$ is from
\ SLNT [6] and $E_{exact}$ from direct numerical integration [5].

\bigskip

\begin{tabular}{|c|c|c|c|c|}
\hline
& $\alpha $ & $2s$ & $3p$ & $4d$ \\ \hline
\multicolumn{1}{|l|}{$-E_{PSLET}$} &  & $0.121234415$ & $0.055496250$ & $%
0.031244807$ \\ 
\multicolumn{1}{|l|}{$-E_{44}$} & $0.1$ & $0.126937229$ & $0.055498046$ & $%
0.031244799$ \\ 
\multicolumn{1}{|l|}{$-E_{SLNT}$} & $-$ & $-$ & $-$ & $-$ \\ 
\multicolumn{1}{|l|}{$-E_{exact}$} &  & \multicolumn{1}{|l|}{$0.12182090$} & 
\multicolumn{1}{|l|}{$0.05549523$} & \multicolumn{1}{|l|}{$0.03124480$} \\ 
\hline\hline
&  & $0.089679150$ & $0.052114869$ & $0.030781155$ \\ 
& $1.0$ & $0.095048845$ & $0.052110018$ & $0.030781125$ \\ 
&  & \multicolumn{1}{|l|}{$0.089903$} & \multicolumn{1}{|l|}{$0.052509$} & 
\multicolumn{1}{|l|}{$0.030838$} \\ 
&  & \multicolumn{1}{|l|}{$0.09267933$} & \multicolumn{1}{|l|}{$0.05206038$}
& \multicolumn{1}{|l|}{$0.03078150$} \\ \hline\hline
&  & $0.037161915$ & $0.028313951$ & $0.021412017$ \\ 
& $10$ & $0.037154303$ & $0.028313599$ & $0.021412537$ \\ 
&  & \multicolumn{1}{|l|}{$0.037111$} & \multicolumn{1}{|l|}{$0.028225$} & 
\multicolumn{1}{|l|}{$0.021353$} \\ 
&  & \multicolumn{1}{|l|}{$0.03715440$} & \multicolumn{1}{|l|}{$0.02831369$}
& \multicolumn{1}{|l|}{$0.02141257$} \\ \hline\hline
&  & $0.012450958$ & $0.010878999$ & $0.009476019$ \\ 
& $50$ & $0.012450960$ & $0.010879000$ & $0.009476018$ \\ 
&  & \multicolumn{1}{|l|}{$0.016263$} & \multicolumn{1}{|l|}{$0.010882$} & 
\multicolumn{1}{|l|}{$0.009477$} \\ 
&  & \multicolumn{1}{|l|}{$0.01245096$} & \multicolumn{1}{|l|}{$0.01087900$}
& \multicolumn{1}{|l|}{$0.00947602$} \\ \hline\hline
&  & $0.003923030$ & $0.003659489$ & $0.003410238$ \\ 
& $200$ & $0.003923030$ & $0.003659489$ & $0.003410238$ \\ 
&  & \multicolumn{1}{|l|}{$0.004503$} & \multicolumn{1}{|l|}{$0.003660$} & 
\multicolumn{1}{|l|}{$0.003411$} \\ 
&  & \multicolumn{1}{|l|}{$0.00392303$} & \multicolumn{1}{|l|}{$0.00365949$}
& \multicolumn{1}{|l|}{$0.00341024$} \\ \hline
\end{tabular}

\newpage

\begin{center}
\bigskip \FRAME{itbpFU}{175.125pt}{175.125pt}{0pt}{\Qcb{(a)}}{\Qlb{(a)}}{%
Figure }{\special{language "Scientific Word";type
"GRAPHIC";maintain-aspect-ratio TRUE;display "PICT";valid_file "T";width
175.125pt;height 175.125pt;depth 0pt;original-width 63.25pt;original-height
63.25pt;cropleft "0";croptop "1";cropright "1";cropbottom "0";tempfilename
'FRCU0200.wmf';tempfile-properties "XPR";}} \ \ \ \ \FRAME{itbpFU}{173.625pt%
}{173.625pt}{0pt}{\Qcb{(b)}}{\Qlb{(b)}}{Figure }{\special{language
"Scientific Word";type "GRAPHIC";maintain-aspect-ratio TRUE;display
"PICT";valid_file "T";width 173.625pt;height 173.625pt;depth
0pt;original-width 63.25pt;original-height 63.25pt;cropleft "0";croptop
"1";cropright "1";cropbottom "0";tempfilename
'FRCUO504.wmf';tempfile-properties "XPR";}} \ \ \\[0pt]
\ \ \ \ \ \FRAME{itbpFU}{184.1875pt}{184.1875pt}{-11.3125pt}{\Qcb{(c)}}{\Qlb{%
(c)}}{Figure }{\special{language "Scientific Word";type
"GRAPHIC";maintain-aspect-ratio TRUE;display "PICT";valid_file "T";width
184.1875pt;height 184.1875pt;depth -11.3125pt;original-width
63.25pt;original-height 63.25pt;cropleft "0";croptop "1";cropright
"1";cropbottom "0";tempfilename 'FRCUWZ06.wmf';tempfile-properties "XPR";}}
\end{center}

Figure(1): The effect of the truncation parameter on the trend of
convergence of Eq.(47) for 1s-state of Eq.(2): (a)$\alpha =0.1,$ (b) $\alpha
=50$, (c) $\alpha =200.$ Where the numbers on the horizontal axis represent
the number of corrections added to the leading energy term of PSLET,
Eq.(47), the vertical axis represents the energies ( in 
h\hskip-.2em\llap{\protect\rule[1.1ex]{.325em}{.1ex}}\hskip.2em%
$=m=1$ units ) and the horizontal curve denotes the exact numerical results
from [5].\ \ \newpage

\begin{center}
\ \FRAME{itbpFU}{184.1875pt}{184.1875pt}{-4.5pt}{\Qcb{(a)}}{}{Figure }{%
\special{language "Scientific Word";type "GRAPHIC";maintain-aspect-ratio
TRUE;display "PICT";valid_file "T";width 184.1875pt;height 184.1875pt;depth
-4.5pt;original-width 63.25pt;original-height 63.25pt;cropleft "0";croptop
"1";cropright "1";cropbottom "0";tempfilename
'FRD2F007.wmf';tempfile-properties "XPR";}}\ \FRAME{itbpFU}{178.8125pt}{%
184.0625pt}{-4.5pt}{\Qcb{(b)}}{\Qlb{(d)}}{Figure }{\special{language
"Scientific Word";type "GRAPHIC";display "PICT";valid_file "T";width
178.8125pt;height 184.0625pt;depth -4.5pt;original-width
63.25pt;original-height 63.25pt;cropleft "0";croptop "0.9994";cropright
"0.9779";cropbottom "0";tempfilename 'FRCUIH03.wmf';tempfile-properties
"XPR";}}

\FRAME{itbpFU}{182.6875pt}{182.6875pt}{0pt}{\Qcb{(c)}}{\Qlb{(e)}}{Figure }{%
\special{language "Scientific Word";type "GRAPHIC";maintain-aspect-ratio
TRUE;display "PICT";valid_file "T";width 182.6875pt;height 182.6875pt;depth
0pt;original-width 63.25pt;original-height 63.25pt;cropleft "0";croptop
"1";cropright "1";cropbottom "0";tempfilename
'FRCUR705.wmf';tempfile-properties "XPR";}}
\end{center}

\bigskip Figure(2): The effect of the angular momentum quantum number, $l,$
and the radial quantum number, $n_{r}$, on the convergence trend of Eq.(47)
for (2) with $\alpha =50$: (a) 1s-state , (b) 3p-state , (c) 4d-state. The
figures prescription is similar to figure 1.

\end{document}